Conductance-Phase Determination in Double Slit Transmission across a Quantum Dot by Hilbert Transform Method

**R. Englman\* and A. Yahalom#**

\*Department of Physics and Applied Mathematics,
Soreq NRC, Yavne 81800 and Research Institute, College of
Judea and Samaria, Ariel 44284, Israel
#Faculty of Engineering. Tel Aviv University,
Ramat Aviv, Israel

(email:englman@vms.huji.ac.il)

.

Recent novel mesoscopic two-arm experiments involving quantum dots, electron interferometry and Aharononov-Bohm effects have enabled measuring the electron transmission probabilities and the phases. Unexpected features in the phases as function of the gap voltage U have simulated intensive theoretical works. It is shown in this paper that the phases (f) and conductances ($|C|$), appearing in both the experimental and the theoretical works, are interrelated through integral expressions, causing f and $\log(|C|)$ to be Hilbert transforms. The empirically found interrelations imply remarkable analytical properties of the U-dependence of wave-functions in mesoscopic systems.

1. Introduction

A short while ago the conductivity and phase of transmitted electrons were jointly determined by employment of a double-slit arrangement in conjunction with a quantum dot[1-3]. The role of the latter was, when inserted in one arm of the interferometer, to induce a measurable phase shift in the component of the electron wave-function representing passage along that arm. In the experiments the gate voltage U was varied (keeping other essential parameters fixed) and the transmission amplitude modulus or some related quantity (named here for short "conductance" and denoted by $|C|$), and phase f were determined as the system passed through several resonances. Surprise was occasioned by the more recent experiments in which the phase increased by $\pi$ across the resonances (manifested by maxima in the transmission amplitude) but dropped sharply by the same amount at anti-resonances [4]. Several theoretical treatments have been published to account for these results [5-9] and further efforts of interpretation are in progress [10, 11]. These works employ a variety of physical models and compare them to the conductance and phase curves.

In contrast the these works, the present Note points out a consistency relationship between the observed phases and transmission amplitudes, which, whenever it holds, is model independent and is grounded in the analyticity properties of the wave transmission. The relation makes $\log|C|$ and f Hilbert transforms ([12], explicitly shown in Eqs.(3) and (4), below) as functions of the externally applied gate voltage parameter. Hints, as to the existence of such relationship can indeed be found in several of the previously quoted papers, (and indeed the Breit-Wigner formula is a particular instance of the applicability of Hilbert transform), but neither the precise form (e.g. the relevance of the <u>logarithm</u> of the conductance), nor the conditions for the validity of the theory have been stated. We shall state these conditions in a later section with the aim of channeling future theoretical attempts to building physical models so that the conditions are explicitly taken care of.

The present work is an outgrowth of previous publications in which reciprocal relations between the phase and moduli of a time dependent wave function and of optical wave fields were obtained and applied [13-17]. These relations operated in the time domain. Reciprocal relations in the frequency domain, taking the form of Kramers-Kronig relations, have of course been widely known [18]. These are based on the causality principle and (although mathematically similar) are logically unrelated to those in [13-17]. The present work extends the formalism to the consideration of the analytical properties of the wave function in a mesoscpic system as function of an external parameter, the gate voltage.

2. Complex conductance.

We consider solutions of a Schrodinger equation or the corresponding propagator or Green function. The Schrodinger equation contains the gate voltage U and therefore the solutions will be functions of U. The same will be true for a generic quantity C(U) that derives from the solutions, like the conductance, the transmission amplitude or "the interference term ". The various "|C|-equivalent" quantities are listed in Table 1, with sources given. We now make the supposition (presently to be confirmed by the observed data) that the conductance |C| is the modulus of a complex quantity

$$C=C(U) \qquad (1)$$

which depends on the real variable U. To complete the definition of C we introduce the phase f, as

$$C(U) = |C(U)| \exp[if(U)] \tag{2}$$

Under certain conditions, the following relations hold between the phase and the conductance:

$$-(1/\pi) P \int_{-\infty}^{\infty} dU' [\log|C(U')|]/(U'-U) = \pm f(U) \tag{3}$$

and

$$(1/\pi) P \int_{-\infty}^{\infty} dU' [f(U')]/(U'-U) = \pm \log|C(U)| \tag{4}$$

Here P denotes the principal value of the integral. $\log|C(U)|$ and $\pm f(U)$ are "Hilbert transforms" [12]. We now turn to the conditions for the validity of (3) and (4).

Let us assume that C can be analytically continued to be a function of the complex "gate voltage"

$$W = U + iV \tag{5}$$

in the sense that if W is inserted in the Schrodinger equation then the solutions reduce to the physical solutions when V --> 0. We thus define C as a complex function of the complex variable W

$$C = C(W) \tag{6}$$

C(W) is now supposed (a) to be analytic in the lower (or upper) half W-plane, (b) to tend to zero on a large semicircle in that half plane and (c) not to have zeros in the half-plane (though it may have zeros on the real-line V=0 [12,14,17]). The lower (or upper) signs are appropriate for functions analytic in the lower (or upper) halves. Certain refinements to these conditions exist. (Thus, in (4) the phase has the freedom of choice of an additive constant and in (3) the conductance that of a multiplicative constant, since the Hilbert transform of a constant is zero. This permits us to treat on the same footing other quantities related to the conductance, as long as they differ from it only through a multiplicative constant. Moreover, when the physical quantities are some powers of each other, then the corresponding derived phases are simply multiples of each other so that if the unit of the phase is not specified, then the relation in (3) can be used for all of them. With this understanding, our results hold equally for conductance, Aharonov-Bohm oscillation amplitudes, transmission probabilities and other quantities. As already stated, we refer to them generically as "Conductance". )

It is clear from (3) and (4) below that (provided the stated conditions, (a) – (c), hold) the phase is uniquely given from the conductivity and vice versa. Any physical model or theory needs to account of one type of quantity alone.

In the following figures we present graphically several types of phases and conductance amplitudes (not the logarithms) as functions of the real gate voltage U and relate them to published experimental and theoretical results. The quantities plotted by us all satisfy Eqs. (3) and (4) and are Hilbert transforms in the U (or W) domain. The actual expressions on which the plots are based are listed in Table 2. We can now add:

Any (observed) conductance that, as function of (real) U, is numerically similar to any of the $|C(U)|$'s in the list and has the same analytical behavior for (complex) W, will also yield a corresponding phase f(U) that is numerically similar. Any conductance that is numerically similar, but is analytically dissimilar, will yield a phase that is completely dissimilar. These properties are reflections of the fact that the conductance derives from a (differential) equation, that is defined for complex values of the gate voltage.

3. Graphical representations.

(Figure 1.)

(Table1.)

The curves in Figure 1 have the shapes of the experimental values of Schuster et al. [4] shown in their Figure 3b and 3c, (or Fig. 2 in [8]) except that the experimental values are somewhat skewed and not quite periodic. The latter property (if not an instrumental effect) can have its origin in differences between levels of the quantum dots. Since the effects appear to be small, we ignore them in this work.

The strong antiresonances near integral multiple values of $\pi$, where $C(U) \approx 0$, and the sudden "phase lapses [8]" there, between resonances are evident.

(Figure 2.)

The curves in Fig. 2 differ from the previous set only by allowing skewness in the conductances, present in the observational curves of [4]. The phase does not significantly differ from that in Fig. 1.

(Figure 3.)

The conductance curves of Fig. 3 are still oscillatory, but they do not get close to the horizontal axis, i.e. $|C(U)|>>0$. Yet, the phases oscillate, contrary to what might have been anticipated. (Note, e.g., the caption to Fig.2 in [9].) The downward slope of the phase is now moderate and, in fact, it scales with $[C(\pi)]^{-1}$.

(Figure 4)

The simple curves in Fig. 4 resemble some experimental and theoretical curves (e.g. Fig. 4(a) and (c) in[1], Fig. 2c in [4], Fig. 1 in [6]), showing that the mathematical relations in (3) and (4) hold between the observational quantities.

(Figure 5.)

The curves in Fig. 5 resemble the theoretical curves of [ 8] (Fig. 3). One notes that the phase, as shown, is fully continuous and does not make an unphysical jump near $U =$ multiple of $2\pi$. (A jump is conventionally introduced to keep arctan uniquely defined. Here we demand the phase to be continuous.) The physical phase that is shown is not "of bounded variation" and therefore Eq. (4) cannot be directly used. Instead, one has first to subtract from the phase a term linear in U. It is the phase thus obtained (that is of proper behavior) which must be used in (4), and also it is this phase which is (by consistency) given by (3). To regain the physical phase, one must then reinstate the linear term. (This procedure is equivalent to the subtraction of the dynamic phase in adiabatic time dependent wave functions to obtain a topological "connection" that is integrable [13].).

4. Hilbert transform for raw data values.

In this section we derive the phase directly from the raw observational data of the magnitudes ("conductance") by the Hilbert transform method (that is, without any interpolating function). Specifically, we start with <u>discrete</u> data values shown by dots in Fig. 3b of [4] and employ Eq. (3) on these. A slight problem arises, though, in that the range of integration in (3) is infinite, while the data points cover only a finite range of the gate voltage. A natural (but perhaps oversimplified) solution of this problem is to assume that the experimental data posses a periodicity (implying, as we have done before, that deviations from strict periodicity are neglilible). We have taken as basis the "conductance" (more accurately: "the magnitude of the Aharonov-Bohm oscillations") data values in Fig 3b of [4] contained in the resonance peak just following the vertical dotted line (since these seem to be the least affected by experimental errors) and used these experimental points adjusted to periodicity. They are shown in Fig. 6A by dots. We have then replaced the infinite integral in Eq. (3) (with the positive sign) by a sum over the experimental values inside the elementary resonance peak and by a further discrete sum over all equivalent, identical peaks. The values for the phases that are thus obtained from the integral in (3) are shown in Fig. 6B by stars. (Fig. 6)

These are in reasonably close agreement with the experimental values of the phase, also given by Heiblum and coworkers [4] in their Fig.3c and shown by us in Fig. 6B by dots (again imposing a periodic recurrence of the peak). The only adjustment that was made in the calculated phase (plotted in units of $\pi$) is a constant vertical shift. (Note our previous remarks about an arbitrariness of a constant shift in the phase in section 2.) The range of the computed phase exceeds the observed one by about 15%: this excess comes presumably by the inaccuracy involved in replacing the principal integral by a sum over the data points (comprising only 16 values). This is probably also the source of the discrepancy near odd-integral values of U. However, the calulated "phase-lapses" are similar to the observed ones and the displacements in the maxima between the conductance and the phase are ¼ of the fundamental period, as given by experiment.

Considering that the Hilbert transform method is based on continuous functions, it is gratifying to note its applicability to discrete, raw, numerical data.

5. Discussion.

In this work relations have been given between observed phases and transmission amplitudes, so that the phase and the log of the "conductance" are Hilbert transforms as function of the gap voltage. The relations are contingent (to certain analyticity conditions), rather than absolute. We have found however, that to a good accuracy available experimental and theoretical curves obey the relationships. This leads to the tentative conclusion that the analyticity requirements (a) –(c) listed in Section 2 hold true and may be indeed be a requisite component of the physical situation.

The relations do not replace a physical model, but provide a check on it. The transmission amplitude modulus, shown in Eq.(1) in [3] and based on simplified one-dimensional models of [14,15], is derivable from a complex conductance

$$C(\varphi) \propto [1-r_1 r_2 \exp(i\varphi)]^{-1} \qquad (7)$$

where $r_1$, $r_2$ are reflectivities, and $\varphi$ is the sum of a magnetic flux term and the phase acquired by the partial wave traveling along the ring's arm in the absence of a magnetic field [15]. The latter part is an essentially linear function of the gate voltage U (Fig. 2(c) in [3]). Since $|r_1 r_2| < 1$, this approximate expression has the postulated analytical behavior in U. Similarly, the transmission amplitude, Eq. (1) in [9], is essentially the difference of two terms of the form in Eq.. (7), with $r_1 r_2$ taking different signs; this, again has the analyticity properties (a)-(c).

In previous works [16-19] in which time was the independent variable, it has been <u>established</u> that in several physically significant cases the analyticity conditions are met. Thus a proof has been given for the ground state of an adiabatically evolving system, including the location of the zeros of the wave-function.[17]. (The same form of analyticity is also present in coherent and squeezed wave packet states.). Can we make similar *a priori* assertions for the wave function of a mesoscopic system as function of an external parameter, like U? To answer this, one notes that, if U appears somehow in the potential of Schrodinger equation, the solution (and quantities derived from it) will not diverge unless the potential does so [21]. However, the Hilbert transform formulation requires also that there should be no zeros in one half-plane of the complex parameter. We have no formal proof that this must generally hold true, but are encouraged by the (approximate) expressions cited above and by agreements found in this work, under assumption of the analyticity conditions.

On the other hand, it may happen in some cases that there are discrepancies from the integral relations. The physical meaning of these deviations might be of interest. Let it be also remarked that a correction term is available for those cases that have extra zeros in the wrong half-plane [13,22]. The contribution from this term is of a fixed sign (and if the zeros are far from the real U-axis, their effect is small) [13].

In conclusion, future experimental or theoretical work on electron transmission in mesoscopic systems should take account of the Hilbert transform relationships between phase and transmission probability. From a broader view, this case appears to be a remarkable, and possibly first, instance in which analytical properties in an "external parameter (U) space" have observable effects.

Acknowledgement.

Our thanks to David Sprinzak for explaining the experiments of the Heiblum group and to Yuval Gefen for discussing the theory.

| "Modulus" quantity | Reference |
|---|---|
| Conductance | [1], [7] |
| Transmission Amplitude | [4], [8] |
| Transmission Coefficient | [3], [9] |
| Amplitude of Aharonov-Bohm Oscillations | [4], [6] |
| Interference Term Amplitude | [8] |

Table1. Physical quantities represented in this work by $|C|$.

| Figure | C(U) |
|---|---|
| 1 | $(1 + .95e^{iU})/(1 - .4e^{iU})$ |
| 2 | $1.01(1+.95e^{iU})/[1-.4(1+.75\sin U)e^{iU}]$ |
| 3 | $(1+.7e^{iU})/(1-.4e^{iU})$ |
| 4 | $[(U-3)+i\sqrt{2}]^{-1}$ |
| 5 | $[1-.8e^{iU}]^{-1}$ |

Table 2. Sources of the plots in Figures 1-5.

In each case $|C(U)|$ and arg $C(U)$ was plotted. Log $|C(U)|$ and arg $C(U)$ (=f) are Hilbert Transforms.

Figure Captions.

Fig, 1 Symmetric, periodic resonances conductances (or Transmission probability amplitudes, etc.) and phases are plotted against the gap voltage U (all in arbitrary units). The source of the plots in this and the following figures are shown in Table 1.

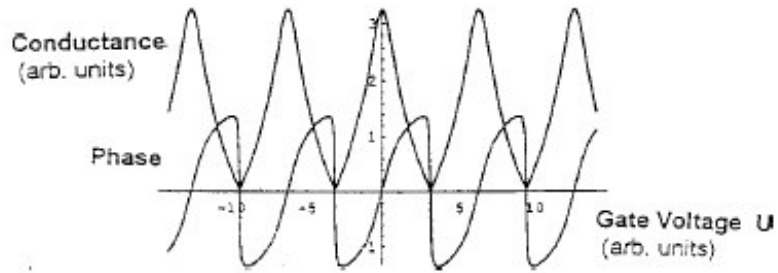

Figure 2. Skewness effects

(Quantities and units as in Fig.1)

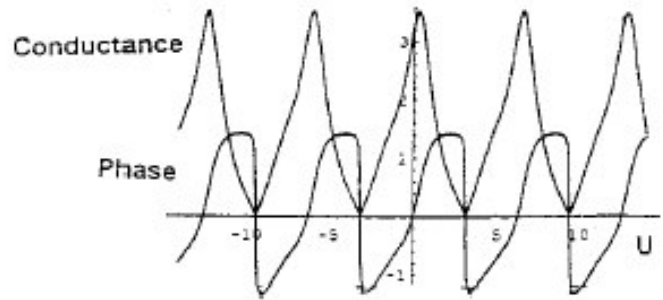

Figure 3  Conductance not having nodes.

(Quantities and units as in Fig.1)

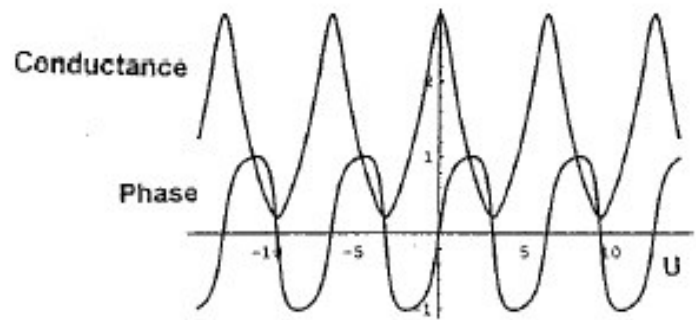

Figure 4. Lorentzian conductance

(Quantities and units as in Fig.1)

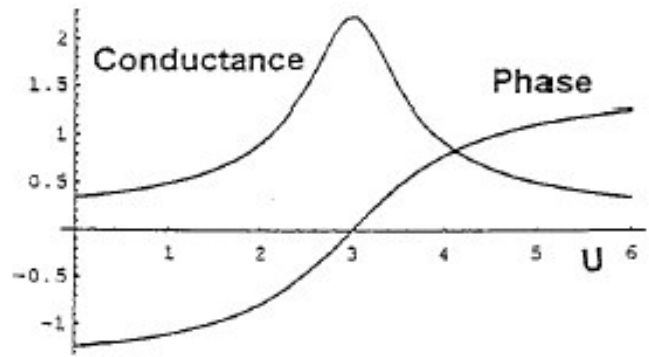

Fig. 5. Stepwise phases.

(Quantities and units as in Fig.1)

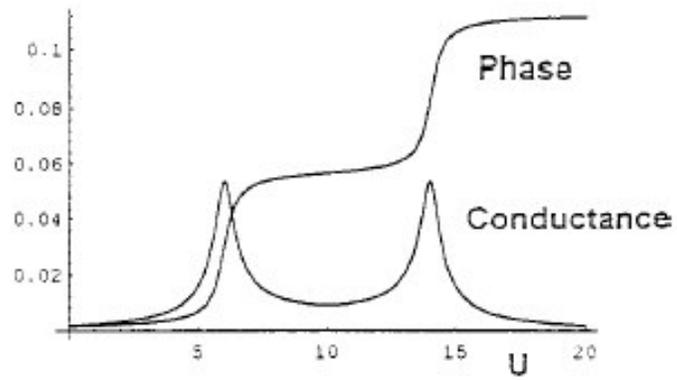

Fig.6. Discrete application of Hilbert transform.

A. Observed oscillation magnitudes. The values shown by dots are from [4], here designated as "Conductance", in arbitrary units.

B. Phase.Angle (in units of π). The stars show the values for f(U) from expression (3), after adding a uniform upward shift of approximately π/2. The dots show values from [4].

Broken lines connect values.

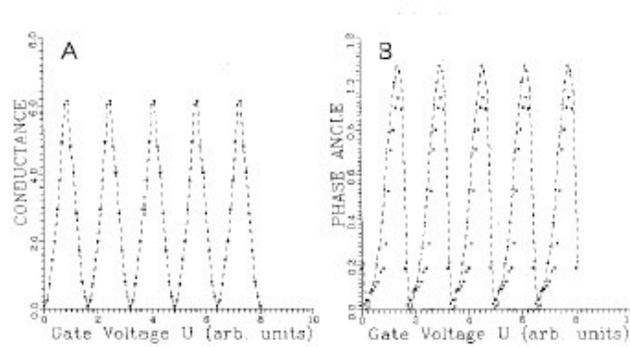